\documentclass[12pt]{article}
\usepackage[a4paper]{geometry}
\usepackage[parfill]{parskip}    % begin paragraphs with an empty line rather than an indent
\usepackage[usenames,dvipsnames]{xcolor}
\usepackage[english]{babel}
\usepackage{cite,enumerate,enumitem,booktabs,float,graphicx}

\makeatletter
\g@addto@macro\bfseries{\boldmath}
\makeatother

\usepackage{amsmath,amssymb,bm,mathtools,tensor}
\usepackage{authblk}

\usepackage[pdftex]{hyperref}
\definecolor{dark-blue}{rgb}{0.15,0.15,0.4}
\hypersetup{
  colorlinks,
  linkcolor={dark-blue},
  citecolor={blue}
}

\newcommand{\beq}{\begin{equation}}
\newcommand{\eeq}{\end{equation}}

\newcommand{\atilde}{\tilde{a}}

\title{\textbf{Semiclassical cosmic string evaporation}}
\date{}

\author[]{Robert Penna\footnote{pennar@sunypoly.edu}}
\affil[]{
  Department of Mathematics and Physics \\
  SUNY Polytechnic Institute,
  Utica, NY 13502 USA
}

\begin{document}

\maketitle

\thispagestyle{empty}

\begin{abstract}

We describe a new tunneling solution for the decay of a cosmic string into a burst of gravitational waves.
We find the relevant instanton and compute the tunneling rate.  Locally, our solution is just an analytic continuation of the Kerr metric (but there is no black hole in our solution).  An interesting feature of our result is that there is a conical singularity in the initial state but there is no singularity in the final state.  This demonstrates that singularities can disappear in quantum gravity in ways that are impossible in classical gravity.

\end{abstract}

\newpage

Nonperturbative results about quantum gravity are hard to come by.  When they do appear, they are usually worth investigating.  One class of examples we have in mind are the ``bubble of nothing'' solutions \cite{Witten:1981gj,Aharony:2002cx,Horowitz:2002cx,Astorino:2022fge}, which show that simple Kaluza-Klein vacuums are unstable.  Another class of examples are the black hole pair production solutions \cite{Garfinkle:1990eq,Garfinkle:1993xk,Eardley:1995au,Emparan:1995je,Hawking:1995zn,Dowker:1995gb}, which shed light on the role of topology change in quantum gravity and give insight into the nature of black hole entropy.  

These are all examples of instanton-mediated tunneling processes.  An instanton is a classical solution of the Euclidean signature equations of motion.  It interpolates between two different field configurations (an initial state in the far Euclidean past and a final state on a surface of fixed Euclidean time).  The action of the instanton gives the semiclassical approximation to the tunneling rate.

In this note, we describe a new example of an instanton-mediated tunneling process in semiclassical gravity.  In our solution, a cosmic string spontaneously decays into a burst of gravitational radiation.  The cosmic string in the initial state is an infinitely thin thread.  The metric has a conical singularity at the location of the string.  The gravitational radiation in the final state, on the other hand, has no singularity.  So our solution gives a clean demonstration that singularities can disappear in quantum gravity in ways that are impossible in classical gravity. 

An interesting feature of our solution is that, locally, it is just an analytic continuation of the Kerr metric\footnote{Different analytic continuations of the Kerr metric give different instantons. For example, there is an analytic continuation which is different from the one considered here which gives a bubble of nothing \cite{Aharony:2002cx,Horowitz:2002cx}.}.  Despite this, the physical interpretation is very different (there is no black hole).  Although analytic continuation is used to get the solution, the metric is real valued.  Our metric differs from Kerr globally because our angular coordinates have different periodicities.  The Lorentzian endpoint of our solution was discovered long ago by Piran, Safier and Katz \cite{Piran:1986fa}.

The gravitational wave burst in the final state has a pulse-shaped profile with a characteristic width, $\atilde$.  The tunneling rate we compute scales as 
\beq
e^{ - (\cdots) \atilde} \,.
\eeq
In other words: the most narrow, sharply peaked bursts are favored all the way down to $\atilde=0$.  In reality, we suspect this an artifact of the thin string approximation.  If we give the string a finite thickness, the tunneling rate will probably turn over and have a maximum at a scale set by the string thickness.  It would be interesting to try to check this.  

To get a nonzero tunneling rate, we need to make the spatial direction along the string periodic.  Let $2\pi R$ be the length of the string.  Our result for the tunneling rate is
\beq\label{eq:rate}
e^{-  \pi \atilde R (\alpha - 1)} \,,
\eeq
where $\alpha=1/(4\mu)$ and $\mu$ is the energy per unit length of the string. The fact that the rate vanishes in the $R\rightarrow \infty$ limit harmonizes well with bubble of nothing calculations.   There too, the rate of decay vanishes in the limit in which the size of the Kaluza-Klein circle becomes infinite.  To make closer contact with Kaluza-Klein bubbles of nothing, it would be interesting to look for a generalization of our solution in five spacetime dimensions\footnote{There is a five dimensional analogue of the Kerr metric called the Myers-Perry metric \cite{Myers:1986un}.  It would be interesting to look for a five dimensional generalization of our solution based on the Myers-Perry metric.  Actually, there already is a well-known instanton based on Myers-Perry but the physical interpretation is very different \cite{Dowker:1995gb}.  In the known construction, the time direction of the instanton comes from analytically continuing the Myers-Perry $\psi$ coordinate.  In our solution, on the other hand, the time direction of the instanton comes from analytically continuing the Weyl $z$ coordinate of the Kerr metric in Weyl coordinates.  So a good starting point for finding a five dimensional analogue of our solution would be to consider the Myers-Perry metric in Weyl coordinates \cite{Harmark:2004rm}.}.  Then the spatial direction along the string could be identified with the fifth dimension of the Kaluza-Klein model.  

The string in the initial state of our solution is static but astrophysical cosmic strings (if they exist at all, or once existed) are not be expected to be static.  Nonstatic strings can radiate through classical channels \cite{Vachaspati:2015cma}.  We expect these conventional modes of radiation to swamp any contribution from our semiclassical mechanism.  We do not anticipate our tunneling solution to be relevant for astrophysical strings.

We have already mentioned that our solution comes from an analytic continuation of the Kerr metric.  It is interesting to match parameters.  Our width parameter, $\atilde$, is related to the spin parameter of the Kerr black hole by $a=i\atilde$.  In other words, we are studying the Kerr metric at pure imaginary values of the spin parameter.  The string's line energy density, $\mu$, is related to the mass and spin parameters of Kerr by
\beq\label{eq:mu}
\mu = \frac{\atilde}{4\sqrt{M^2 + \atilde^2}} \,.
\eeq
It is amusing to note that $M=0$ corresponds to $\mu = 1/4$ (in units with $c=G=1$).  It has been conjectured that this value of $\mu$ is a fundamental upper limit on the energy density of strings set by general relativity \cite{Gibbons:2002iv}.  Eqn. \eqref{eq:mu} shows that the ``maximum tension principle'' of the string is mapped to the zero mass limit of the black hole by analytic continuation.  It is unclear if this fact has deeper significance or if it is just a curiosity.

Our string has a third parameter, $R$, which is absent in the Kerr black hole solution. The reason for this is the following.  $2\pi R$  is the length of the direction along the string.    Under the analytic continuation we have referred to, this spatial circle is mapped to the imaginary time direction of the Kerr black hole.  The latter has a well known periodicity related to the temperature of the black hole and it is completely fixed in terms of $M$ and $a$.  The reason it is fixed is that there is a regularity condition at the event horizon. Our solution has no horizon and thus no regularity condition to enforce.  So the period of the circle enters as a third, independent parameter.

We are now ready to discuss the calculation.  Our starting point is the metric of the eternal cosmic string, 
\beq\label{eq:string}
ds^2 = \alpha^2 (-dt^2+dr^2) + dz^2	+ r^2 d\phi^2 \,,
\eeq
where $1 < \alpha < \infty$ is a constant.  This is almost the metric of Minkowski spacetime in cylindrical coordinates but there are two important differences.  First, the $z$ direction is taken to be periodic with length given by $2\pi R$.  Second, the $\phi$ coordinate runs from $0$ to $2\pi$.  Thanks to the factor of $\alpha$ in the metric, this means the spacetime is asymptotically conical and it has a conical singularity at $r=0$.   The angle subtended by the cone is $2\pi (1 - \alpha^{-1})$.  The above metric is well known to describe the spacetime of a thin cosmic string with line energy density $\mu = 1/(4\alpha)$ \cite{Vilenkin:1981zs}.  

The next step is to analytically continue $t=i\tau$ and thus get
\beq\label{eq:euclideanstring}
ds^2 = \alpha^2 (d\tau^2+dr^2) + dz^2	+ r^2 d\phi^2 \,.
\eeq
This is the Euclidean version of the eternal cosmic string.  An instanton which interpolates between this metric and some other state should  asymptote to this metric in the far Euclidean past $(\tau \rightarrow - \infty)$.  It should also have the same asymptotic structure as eqn. \eqref{eq:euclideanstring} at large $r$.  This motivates us to look for Euclidean solutions of the vacuum field equations with the Weyl-like form
\beq\label{eq:weyl}
ds^2 = e^{2(\gamma -\psi)} (d\tau^2+dr^2) + e^{2\psi} (dz + \omega d\phi)^2	
		+ e^{-2\psi} r^2 d\phi^2 \,.
\eeq
This ansatz reduces to eqn. \eqref{eq:euclideanstring} upon setting $\psi = \omega = 0$ and $e^{2\gamma} = \alpha^2$.  The functions, $\psi,\omega$, and $\gamma$, are assumed to be functions of $\tau$ and $r$ only. 

Actually, we do not have to look far because the analytic continuation of a metric discussed long ago by Piran, Safier, and Katz \cite{Piran:1986fa} does the job perfectly.  Their metric is itself an analytic continuation of the Kerr metric.  
What we obtain is a one parameter family of solutions labeled by $\atilde$ (we are thinking of $\alpha$ and $R$ as fixed properties of the boundary conditions now).  To describe these  solutions, 
define $u=i\tau - r$ and $v = i\tau + r$.  Further define
\begin{align}
\lambda_u &= \atilde^{-1} \left[ (\atilde^2 + u^2)^{1/2} - u \right] , \\
\lambda_v &= \atilde^{-1} \left[ (\atilde^2 + v^2)^{1/2} + v \right] ,
\end{align}
and let $\Xi = 1+ \lambda_u \lambda_v + 2[(1-\alpha^{-2} \lambda_u \lambda_v)]$.   Then the solution we seek is
\begin{align}
e^{2\gamma}	&= 1 + \frac{(\alpha^2-1) (1-\lambda_u \lambda_v)^2}{(1+\lambda_u^2)(1+\lambda_v^2)} \,, \label{eq:gamma} \\
e^{2\psi}		&= \frac{\alpha^2 (1-\lambda_u \lambda_v)^2 + (\lambda_u + \lambda_v)^2}
					{\alpha^2 \Xi^2 +(\lambda_u - \lambda_v)^2} \,,	\label{eq:psi} \\
\omega		&= - \atilde (\alpha^2 -1)^{1/2} 
					\frac{\Xi (\lambda_u \lambda_v)^{-1/2} (\lambda_u+\lambda_v)^2}
					{ \alpha^2 (1-\lambda_u \lambda_v)^2 + (\lambda_u + \lambda_v)^2 } \,. \label{eq:omega}
\end{align}
It is straightforward to check that this metric reduces to eqn. \eqref{eq:euclideanstring} in the far Euclidean past and it has the same asymptotic structure as the eternal string at large $r$.

The functions quoted above are not particularly illuminating.  To understand this space, observe that the metric functions are independent of $z$ and $\phi$, so the solution is cylindrically symmetric.  This means we have a conserved quantity, the $C$ energy \cite{Thorne:1965nft}.  A convenient fact about the Weyl-like ansatz \eqref{eq:weyl} is that we can immediately read off this energy: the function $\gamma(\tau,r)$ represents the $C$ energy inside radius $r$ at Euclidean time $\tau$.  The total $C$ energy, $\gamma(\tau,r=\infty) = \log(\alpha)$, is a constant.  It measures the size of the conical deficit at infinity.

\begin{figure}
\begin{center}
\includegraphics[width=0.6\textwidth]{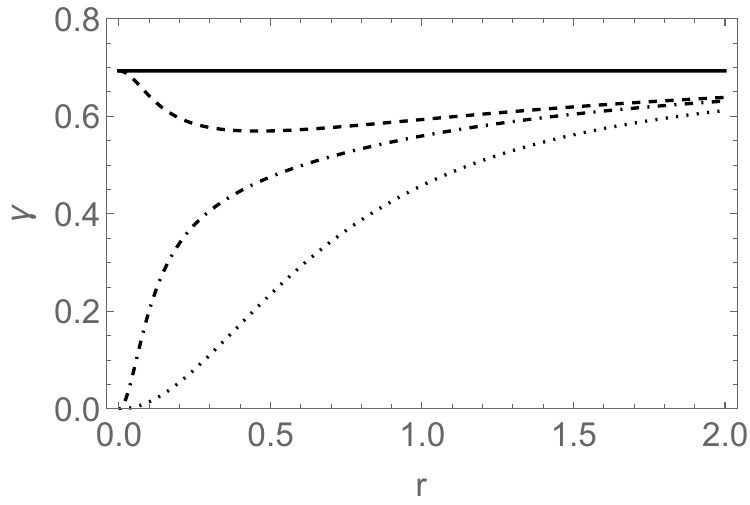}
\caption{Plots of $\gamma(r)$ at Euclidean times $\tau = -20$ (solid), $-1.1$ (dashed), $-0.9$ (dot-dashed), and 0 (dotted).  At large negative $\tau$, all of the $C$ energy is at $r=0$.  The plotted solution has $\atilde=1$ and $\alpha=2$. } 
\label{fig:instanton}
\end{center}
\end{figure}

Something surprising happens if we look at the evolution of $\gamma(r)$ in Euclidean time. Figure \ref{fig:instanton} shows the picture.  At very early times, $\gamma(r) = \log(\alpha)$.  This means all of the $C$ energy is concentrated at $r=0$, which is consistent with the fact that we have a cosmic string in the initial state.  As we approach $\tau=-\atilde$, a negative $C$ energy ``bubble'' appears at small $r$.   This shows up as a dip in the dashed curve of Figure \ref{fig:instanton}.     Negative $C$ energy is impossible in Lorentzian signature, which underscores that this is a classically forbidden process, possible only because we are in Euclidean signature.  

At $\tau = -\atilde$, the value of $\gamma(r)$ at $r=0$ jumps down to zero.  This signals the disappearance of the cosmic string.  After the string disappears, the $C$ energy is everywhere positive.    The total $C$ energy is conserved throughout this process.

What is the endpoint of this transition?  To answer this question, we cut the instanton at $\tau=0$.  This is a moment of time symmetry, so we can match the Euclidean solution to its Lorentzian continuation at this surface without any problems \cite{Gibbons:1990ns}.  The complete real tunneling geometry is defined by evolving the Euclidean solution from $\tau = -\infty$ up to $\tau = 0$, and then continuing forward in Lorentzian time, $t = i \tau$, starting at $t=0$.  

\begin{figure}
\begin{center}
\includegraphics[width=0.4\textwidth]{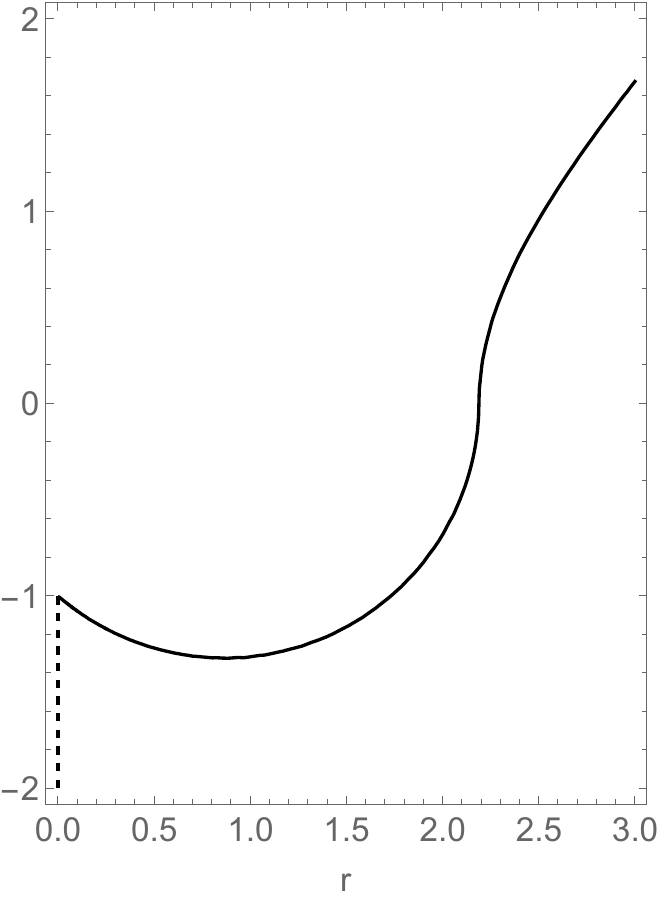}
\caption{The positive vertical axis is Lorentzian time ($t$) and the negative vertical axis is Euclidean time ($\tau$). The dashed line is the string.  The solid black contour indicates the distribution of $C$ energy: $90\%$ of the $C$ energy is to the left of the contour.  The decay of the string is mediated by a negative $C$ energy bubble near $\tau=-1$.  This solution has $\atilde = 1$ and $\alpha = 2$.} 
\label{fig:tunnel}
\end{center}
\end{figure}

The Lorentzian side of the solution has already been studied by Piran, Safier, and Katz \cite{Piran:1986fa}.  For $t>0$, it describes an outgoing burst of gravitational waves.  The metric is smooth at the origin but asymptotically conical.  Figure \ref{fig:tunnel} depicts the distribution of $C$ energy in the full tunneling geometry: $90\%$ of the $C$ energy is to the left of the solid black contour.   The dip near $\tau = -1$ is the negative $C$ energy bubble. The dashed line is the string.  We have arrived at a more or less standard picture of an instanton-mediated tunneling event (compare Figure 2 of \cite{Vilenkin:1983xq}).   The fact that there is a cosmic string in the initial state but not the final state means the initial and final states differ by a superrotation \cite{Strominger:2016wns}.

Finally, we compute the tunneling rate.  The semiclassical tunneling rate is proportional to $e^{-I}$, where $I$ is the action of the Euclidean solution.  
In Euclidean signature, the action is
\beq\label{eq:I}
I = - \frac{1}{16 \pi G} \int_M d^4 x \sqrt{g} \mathcal{R} 
		- \frac{1}{8 \pi G} \int_{\partial M} d^3 x \sqrt{h} K \,.
\eeq
We have included the Gibbons-Hawking boundary term but we will now argue that it is not important for us.  Suppose we cut off the spacetime at a large but finite value of $r = r_c$.  The outward unit normal on the cut off surface is
\beq
n^\alpha = ( 0, e^{-(\gamma-\psi)}, 0, 0 ) \,,
\eeq
evaluated at $r=r_c$.  The extrinsic curvature is $K = {n^\alpha}_{;\alpha}$.  A calculation gives
\beq
\sqrt{h} K = 1 + O(r_c^{-1}) \,.  
\eeq
Naively, this looks like a problem.  Plugging this into the action and integrating over $\tau$ gives infinity.  To regulate this infinity, we need to perform the usual trick of subtracting off a contribution from the  background.  For us, the background is the eternal Euclidean string  \eqref{eq:euclideanstring}.  This background has precisely $\sqrt{h} K = 1$.  So subtracting off the background  completely eliminates the Gibbons-Hawking term.  

Now turn to the Einstein-Hilbert term.  We are studying a vacuum metric with zero cosmological constant so one could be tempted to set $\mathcal{R} = 0$.  But we have to be more careful because there is a conical singularity in the $r$-$\phi$ plane  when $\tau < -\atilde $ and when $\tau > \atilde$.  Suppose we have a conical singularity at a point $P$ in two dimensions which subtends an angle $2\pi + \varepsilon$.  The conical singularity makes a contribution to the scalar curvature given by
\beq
\mathcal{R} = -2\varepsilon \delta_P + \dots
\eeq
The conical singularity in our solution has $\varepsilon = -2\pi(1-\alpha^{-1})$, and $\alpha^{-1} = 4\mu$, so the contribution is $4\pi (1-4\mu)$.  Naively, the integral over $\tau$ is infinite but we regulate it by subtracting off the contribution from the eternal string \eqref{eq:euclideanstring}.   Recall that the string is only present in our solution for $\tau > \atilde$ and $\tau < -\atilde$, so we have the following subtractions:
\beq
\int_{\atilde}^\infty d\tau - \int_0^\infty d\tau = - \int_0^{\atilde} d\tau = -\atilde \,,
\eeq
and 
\beq
\int_{-\infty}^{-\atilde} d\tau - \int_{-\infty}^0 d\tau = -\int_{-\atilde}^0 d\tau = -\atilde \,.
\eeq
So we pick up  a factor of $-2\atilde$ from the $\tau$ integral.  Finally, we have a factor of $2\pi R$ from the $z$ integral and a factor of $\alpha$ from $\sqrt{g}$.  
Putting everything together,
\beq
I = -\frac{1}{16\pi} \alpha \left[ 4\pi (1-4\mu)  \right] (-2\atilde) (2\pi R)  = \pi \atilde R (\alpha - 1) \,.
\eeq
Thus, the end result for the tunneling rate is
\beq
e^{-  \pi \atilde R (\alpha - 1)} \,,
\eeq
which is eqn. \eqref{eq:rate}.   

\subsection*{Acknowledgements}

I am grateful to Roberto Emparan, Gary Horowitz, and Max Niedermaier for comments on the manuscript.

\bibliographystyle{JHEP}
\bibliography{instanton}

\providecommand{\href}[2]{#2}\begingroup\raggedright\begin{thebibliography}{10}

\bibitem{Witten:1981gj}
E.~Witten, \emph{{Instability of the Kaluza-Klein vacuum}},
  \href{https://doi.org/10.1016/0550-3213(82)90007-4}{\emph{Nucl. Phys. B}
  {\bfseries 195} (1982) 481}.

\bibitem{Aharony:2002cx}
O.~Aharony, M.~Fabinger, G.T.~Horowitz and E.~Silverstein, \emph{{Clean time
  dependent string backgrounds from bubble baths}},
  \href{https://doi.org/10.1088/1126-6708/2002/07/007}{\emph{JHEP} {\bfseries
  07} (2002) 007} [\href{https://arxiv.org/abs/hep-th/0204158}{{\ttfamily
  hep-th/0204158}}].

\bibitem{Horowitz:2002cx}
G.T.~Horowitz and K.~Maeda, \emph{{Colliding Kaluza-Klein bubbles}},
  \href{https://doi.org/10.1088/0264-9381/19/21/317}{\emph{Class. Quant. Grav.}
  {\bfseries 19} (2002) 5543}
  [\href{https://arxiv.org/abs/hep-th/0207270}{{\ttfamily hep-th/0207270}}].

\bibitem{Astorino:2022fge}
M.~Astorino, R.~Emparan and A.~Vigan{\`o}, \emph{Bubbles of nothing in binary
  black holes and black rings, and viceversa},
  \href{https://doi.org/10.1007/JHEP07(2022)007}{\emph{JHEP} {\bfseries 07}
  (2022) 007} [\href{https://arxiv.org/abs/2204.09690}{{\ttfamily
  2204.09690}}].

\bibitem{Garfinkle:1990eq}
D.~Garfinkle and A.~Strominger, \emph{{Semiclassical Wheeler wormhole
  production}}, \href{https://doi.org/10.1016/0370-2693(91)90665-D}{\emph{Phys.
  Lett. B} {\bfseries 256} (1991) 146}.

\bibitem{Garfinkle:1993xk}
D.~Garfinkle, S.B.~Giddings and A.~Strominger, \emph{{Entropy in black hole
  pair production}}, \href{https://doi.org/10.1103/PhysRevD.49.958}{\emph{Phys.
  Rev. D} {\bfseries 49} (1994) 958}
  [\href{https://arxiv.org/abs/gr-qc/9306023}{{\ttfamily gr-qc/9306023}}].

\bibitem{Eardley:1995au}
D.M.~Eardley, G.T.~Horowitz, D.A.~Kastor and J.H.~Traschen, \emph{{Breaking
  cosmic strings without monopoles}},
  \href{https://doi.org/10.1103/PhysRevLett.75.3390}{\emph{Phys. Rev. Lett.}
  {\bfseries 75} (1995) 3390}
  [\href{https://arxiv.org/abs/gr-qc/9506041}{{\ttfamily gr-qc/9506041}}].

\bibitem{Emparan:1995je}
R.~Emparan, \emph{{Pair creation of black holes joined by cosmic strings}},
  \href{https://doi.org/10.1103/PhysRevLett.75.3386}{\emph{Phys. Rev. Lett.}
  {\bfseries 75} (1995) 3386}
  [\href{https://arxiv.org/abs/gr-qc/9506025}{{\ttfamily gr-qc/9506025}}].

\bibitem{Hawking:1995zn}
S.W.~Hawking and S.F.~Ross, \emph{{Pair production of black holes on cosmic
  strings}}, \href{https://doi.org/10.1103/PhysRevLett.75.3382}{\emph{Phys.
  Rev. Lett.} {\bfseries 75} (1995) 3382}
  [\href{https://arxiv.org/abs/gr-qc/9506020}{{\ttfamily gr-qc/9506020}}].

\bibitem{Dowker:1995gb}
F.~Dowker, J.P.~Gauntlett, G.W.~Gibbons and G.T.~Horowitz, \emph{The decay of
  magnetic fields in {Kaluza}-{Klein} theory},
  \href{https://doi.org/10.1103/PhysRevD.52.6929}{\emph{Phys. Rev. D}
  {\bfseries 52} (1995) 6929}
  [\href{https://arxiv.org/abs/hep-th/9507143}{{\ttfamily hep-th/9507143}}].

\bibitem{Piran:1986fa}
T.~Piran, P.N.~Safier and J.~Katz, \emph{Cylindrical gravitational waves with
  two degrees of freedom: An exact solution},
  \href{https://doi.org/10.1103/PhysRevD.34.331}{\emph{Phys. Rev. D} {\bfseries
  34} (1986) 331}.

\bibitem{Myers:1986un}
R.C.~Myers and M.J.~Perry, \emph{Black holes in higher dimensional
  space-times},
  \href{https://doi.org/10.1016/0003-4916(86)90186-7}{\emph{Annals Phys.}
  {\bfseries 172} (1986) 304}.

\bibitem{Harmark:2004rm}
T.~Harmark, \emph{Stationary and axisymmetric solutions of higher-dimensional
  general relativity},
  \href{https://doi.org/10.1103/PhysRevD.70.124002}{\emph{Phys. Rev. D}
  {\bfseries 70} (2004) 124002}
  [\href{https://arxiv.org/abs/hep-th/0408141}{{\ttfamily hep-th/0408141}}].

\bibitem{Vachaspati:2015cma}
T.~Vachaspati, L.~Pogosian and D.~Steer, \emph{{Cosmic Strings}},
  \href{https://doi.org/10.4249/scholarpedia.31682}{\emph{Scholarpedia}
  {\bfseries 10} (2015) 31682}
  [\href{https://arxiv.org/abs/1506.04039}{{\ttfamily 1506.04039}}].

\bibitem{Gibbons:2002iv}
G.W.~Gibbons, \emph{The maximum tension principle in general relativity},
  \href{https://doi.org/10.1023/A:1022370717626}{\emph{Found. Phys.} {\bfseries
  32} (2002) 1891} [\href{https://arxiv.org/abs/hep-th/0210109}{{\ttfamily
  hep-th/0210109}}].

\bibitem{Vilenkin:1981zs}
A.~Vilenkin, \emph{Gravitational field of vacuum domain walls and strings},
  \href{https://doi.org/10.1103/PhysRevD.23.852}{\emph{Phys. Rev. D} {\bfseries
  23} (1981) 852}.

\bibitem{Thorne:1965nft}
K.S.~Thorne, \emph{Energy of infinitely long, cylindrically symmetric systems
  in general relativity},
  \href{https://doi.org/10.1103/PhysRev.138.B251}{\emph{Phys. Rev.} {\bfseries
  138} (1965) B251}.

\bibitem{Gibbons:1990ns}
G.W.~Gibbons and J.B.~Hartle, \emph{Real tunneling geometries and the large
  scale topology of the universe},
  \href{https://doi.org/10.1103/PhysRevD.42.2458}{\emph{Phys. Rev. D}
  {\bfseries 42} (1990) 2458}.

\bibitem{Vilenkin:1983xq}
A.~Vilenkin, \emph{The birth of inflationary universes},
  \href{https://doi.org/10.1103/PhysRevD.27.2848}{\emph{Phys. Rev. D}
  {\bfseries 27} (1983) 2848}.

\bibitem{Strominger:2016wns}
A.~Strominger and A.~Zhiboedov, \emph{Superrotations and black hole pair
  creation}, \href{https://doi.org/10.1088/1361-6382/aa5b5f}{\emph{Class.
  Quant. Grav.} {\bfseries 34} (2017) 064002}
  [\href{https://arxiv.org/abs/1610.00639}{{\ttfamily 1610.00639}}].

\end{thebibliography}\endgroup

\end{document}